\documentclass[sigplan,screen,nonacm]{acmart}

\AtBeginDocument{%
  }

\usepackage{algorithm}
\usepackage{algorithmicx}
\usepackage{algpseudocode}

\copyrightyear{2025}
\acmYear{2025}
\setcopyright{cc}
\setcctype{by-nc-sa}
\acmConference[EuroMLSys '25]{The 5th Workshop on Machine Learning and Systems }{March 30-April 3, 2025}{Rotterdam, Netherlands}
\acmBooktitle{The 5th Workshop on Machine Learning and Systems (EuroMLSys '25), March 30-April 3, 2025, Rotterdam, Netherlands}\acmDOI{10.1145/3721146.3721948}
\acmISBN{979-8-4007-1538-9/2025/03}

\begin{document}

\title{Exploiting Unstructured Sparsity in Fully Homomorphic Encrypted DNNs}

\author{Aidan Ferguson$^{1}$, Perry Gibson$^{1}$, Lara D'Agata$^{1}$, Parker McLeod$^{2}$, \\ 
Ferhat Yaman$^{2}$, Amitabh Das$^{2}$, Ian Colbert$^{2}$, Jos\'e Cano$^{1}$}
\affiliation{%
\institution{University of Glasgow, UK$^{1}$ \hspace{0.3em} AMD$^{2}$}
\country{}
\thanks{Perry Gibson contributed to this work while at the University of Glasgow.}}

\renewcommand{\shortauthors}{Ferguson et al.}


\begin{abstract}

The deployment of deep neural networks (DNNs) in privacy-sensitive environments is constrained by computational overheads in fully homomorphic encryption (FHE). 
This paper explores unstructured sparsity in FHE matrix multiplication schemes as a means of reducing this burden while maintaining model accuracy requirements. 
We demonstrate that sparsity can be exploited in arbitrary matrix multiplication, providing runtime benefits compared to a baseline na\"ive algorithm at \emph{all} sparsity levels. This is a notable departure from the plaintext domain, where there is a trade-off between sparsity and the overhead of the sparse multiplication algorithm. 
In addition, we propose three sparse multiplication schemes in FHE based on common plaintext sparse encodings. We demonstrate the performance gain is scheme-invariant; however, some sparse schemes vastly reduce the memory storage requirements of the encrypted matrix at high sparsity values. 
Our proposed sparse schemes yield an average performance gain of $2.5\times$ at $50\%$ unstructured sparsity, with our multi-threading scheme providing a $32.5\times$ performance increase over the equivalent single-threaded sparse computation when utilizing $64$ cores.

\end{abstract}




\maketitle


\section{Introduction}

Deep neural networks (DNNs) have revolutionized the field of artificial intelligence (AI). These models are now integral to many real-world applications, from autonomous driving to healthcare diagnostics~\cite{maugara2021ml, gupta_gupta_shabaz_sharma_2022}. However, utilizing DNNs in privacy-sensitive environments, such as healthcare, presents unique and challenging requirements to process sensitive data while maintaining confidentiality. Traditional encryption mechanisms ensure confidentiality in transit, leaving the underlying data exposed at inference time. 

Fully homomorphic encryption (FHE)~\cite{gentry_2009} has emerged as a powerful cryptographic technique that allows for computation on encrypted data. Despite its promise for privacy-preserving machine learning, FHE incurs a significant computational overhead due to the complexity of its encrypted computations, the growth of noise during operations, and the need for additional procedures to maintain correctness and accuracy for decryption. This overhead often rules out running real-time workloads such as DNN inference, a problem that is emphasized by the trend toward larger, more computationally expensive DNN models~\cite{jiang_zhong_zhou_2023}.

\begin{figure}[t]
    \centering
    \includegraphics[width=0.99\linewidth]{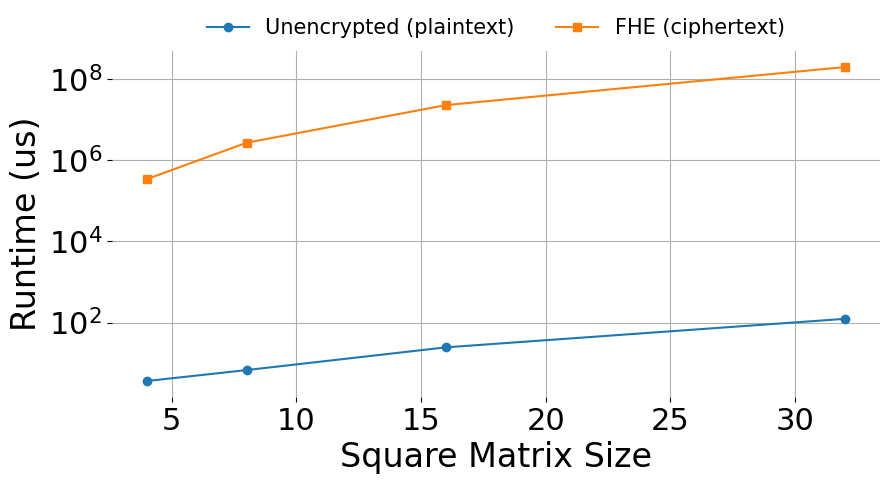}
    \caption{Execution time of square matrix multiplication in plaintext and FHE. Note that the Y-Axis is logarithmic.}
    \label{fig:plaintext-comparison}
\end{figure}

We target the most computationally intensive operation in DNN inference: matrix multiplication (\verb|matmul|)~\cite{haris_SECDA_2021, haris_SECDA-TFLite_2023, transformer_performance, haris_SECDA-LLM_2024}. We observed up to $10^6 \times$ higher runtime when multiplying square matrices in FHE compared to the unencrypted (plaintext) domain (see \autoref{fig:plaintext-comparison}). 
With our proposed sparse multiplication schemes, we show that utilizing unstructured sparsity in arbitrarily sized matrix operands can yield improved execution time in all cases relative to a na\"ive dense implementation; in contrast, plaintext sparsity often requires $\geq 70\%$ to become advantageous due to overheads~\cite{gale_zaharia_brain_elsen_2020}. 
We also propose a multi-threading scheme that exhibits $32.5\times$ performance gain over a single-threaded implementation when utilizing $64$ threads on an AMD EPYC platform. Furthermore, the multi-threaded approach can be applied to arbitrary FHE schemes and implementations.

The contributions of this paper are as follows: i) three unstructured sparse FHE matrix multiplication schemes which provide performance improvements over the na\"ive dense multiplication at all sparsity levels, and operate on arbitrarily sized matrices without structure requirements; and ii) a CPU-based multithreading scheme for sparse matrix multiplication and evaluation on an AMD EPYC CPU. 
We make our C++ implementation with Python bindings available at: \url{https://github.com/aidan-ferguson/sparse-fhe-matmul}.

\section{Background and Related Work}
\label{sec:background}

Previous works that explore dense matrix multiplication within FHE often exploit special matrix structures such as square matrices~\cite{deevashwer} or square binary matrices~\cite{hiromasa_abe_okamoto_2016}. These structures are restrictive for DNN inference, where non-square matrix operations are required. 

Alternative approaches utilize mathematical structures such as hypercubes in the BGV scheme~\cite{huang_zong_2022, deevashwer}, which is inappropriate for DNN inference as discussed in~\autoref{sec:fhe-scheme}. Existing na\"ive matrix multiplication schemes serve as our baseline for comparison with sparse implementations~\cite{cui_chen_lyu_yang_wang_2021}. 

Existing attempts at applying sparsity to the FHE matrix multiplication problem often have restrictions. For example, one such restriction entails only applying sparsity to linear systems of form $Ax = b$, where $A$ is a strictly diagonally dominant matrix~\cite{xiaofeng}. Further work that promises unrestricted sparse matrix multiplication~\cite{chen_zhou, cui_chen_lyu_yang_wang_2021} does not document how the application of sparsity affects runtime. 
Furthermore, these schemes only define dense-sparse matrix multiplication. 
In this work, we focus on exploiting sparsity in both operands, which has direct applications to accelerating DNN architectures with ReLU activation functions that often yield high unstructured activation sparsity~\cite{network_trimming}. Note that, while already common in convolution architectures (e.g., VGGNet~\cite{Simonyan15}, ResNet~\cite{He2015DeepRL}, etc.), ReLU activations are becoming more prominent in foundation LLMs~\cite{relu_strikes_back}.

Finally, open-source dense FHE matrix-multiplication implementations are scarce. TenSEAL~\cite{tenseal2021}, a Python wrapper for Microsoft SEAL~\cite{sealcrypto}, implements dense vector-matrix multiplication. This precludes applications that require full matrix-matrix multiplication. HEMat~\cite{hemat} is the most prominent open-source implementation of dense FHE \verb|matmul| we could find, built using the HEAAN FHE library~\cite{snucrypto_2023}. As such, we utilize HEMat in the process of profiling our results as an additional baseline showing performance without utilizing sparsity. 
To the best of our knowledge, no open-source FHE sparse \verb|matmul| implementations exist; as such we have open-sourced our implementation.

\section{Methodology}


\subsection{FHE Scheme}
\label{sec:fhe-scheme}

We utilize the Microsoft SEAL library~\cite{sealcrypto} which implements three commonly used FHE schemes: Brakerski/Fan-Vercauteren (BFV)~\cite{bfv_citation}, Brakerski-Gentry-Vaikuntanathan (BGV)~\cite{bgv_citation}, and Cheon-Kim-Kim-Song (CKKS)~\cite{cheon_kim_kim_song_2017}. 

CKKS supports approximate floating-point computations and is commonly used in the context of DNN inference~\cite{ckks_justifications, privacy_preserving_ml}, where some approximation and reduced precision are acceptable. 
However, BFV/BGV work only with integer values and do not provide rescaling, which can lead to value growth and potential overflow of plaintext values if the multiplications are not managed properly. This is a key reason why BFV/BGV are less suited for deep computations compared to CKKS, which manages scale growth explicitly through rescaling. BFV/BGV also have limited multiplicative depth and they struggle with DNN computations unless bootstrapping~\footnote{An essential method for restoring ciphertext noise, allowing continuous computation of encrypted data without limitations.} is applied frequently, making CKKS a better choice in this context. 
Furthermore, there are implementations of bootstrapping with the CKKS scheme that execute in more practical time frames than alternative schemes~\cite{alexandru2024general}. Additionally, in CKKS we can utilize polynomial approximations for common DNN activation functions such as ReLU without degrading accuracy~\cite{relu_approximation}.

Motivated by these reasons, we use CKKS as our FHE scheme. The polynomial modulus degree of the scheme is represented by $p$, where $p = 2^{n}$ for some $10 \geq n \geq 15$ and $n \in \mathbb{N}$~\cite{HomomorphicEncryptionSecurityStandard}. Larger values allow for more complex computations at the expense of slower homomorphic operations. We choose $p = 8192$ as the lowest polynomial modulus degree that accommodates our circuit depth. 
For the coefficient modulus, we choose a bit size vector of $\{50, 40, 40, 40, 40\}$ and an initial ciphertext scale of $2^{40}$. These parameters provide enough precision and modulus switches to perform a matrix multiplication followed by an activation function, a single fully connected layer pass-through.


\subsection{Matrix Chunking}

The rotation operation is fundamental in CKKS, performing cyclic rotations of an encrypted vector of length $\frac{p}{2}$, where $p$ is the polynomial modulus degree. Rotations are performed using computationally expensive key-switching operations~\cite{key-switching-costly}, which modify the ciphertext and introduce noise. When many rotations are performed successively, this noise can accumulate instead of padding the ciphertext correctly, leading to errors in the data. This would cause the ciphertext to result in inaccurate values when decrypted.

For this reason, it is preferable to avoid rotations where possible. This can be achieved by implementing a chunking scheme for encoding matrix values into collections of ciphertexts. A chunk size parameter $c$ controls how many values are encoded into each ciphertext, which can each encode a maximum of $\frac{p}{2}$ values. This places an upper bound on the number of rotations performed on each ciphertext to $c$, leading to a more accurate result at the expense of memory. For a non-sparse matrix of size $n \times m$, we must construct $\lceil \frac{n \cdot m}{c} \rceil$ ciphertexts. Furthermore, small chunk sizes facilitate our multi-threading approach by limiting the need for synchronization between threads.


\subsection{FHE Matrix Multiplication Schemes}

\subsubsection{Dense Schemes}
\label{sec:dense-schemes}

Dense schemes that do not exploit any sparsity serve as our baseline for comparison. We utilize two dense schemes: na\"ive and HEMat~\cite{hemat}.
 
\textbf{Na\"ive Dense.} Our baseline implementation demonstrates a na\"ive approach to matrix multiplication, providing a lower performance bound for alternative methods. It does not utilize any sparsity in the matrix. This approach can be seen in Algorithm \autoref{alg:dense-fhe-matmul}. For each result value we select the corresponding row and column vectors from the input matrices, masking the required index and accumulating the value in slot zero of a ciphertext which is then inserted into the encrypted result matrix.

\textbf{HEMat}. We also compare an open-source implementation of HEMat~\cite{hemat_github}, which restricts our evaluation to matrices of dimensions $2^n \times 2^m$. Although our schemes support arbitrarily sized matrices, we conform to this requirement to evaluate against HEMat. HEMat employs the Number Theory Library (NTL) for distributing work across threads. Our multi-threading solution exhibits superior scaling to higher thread counts than HEMat, resulting in different behaviors as the number of threads increases.






\subsubsection{Sparse Schemes}

The encrypted nature of the matrix values means that we cannot perform conditional logic on it. Therefore, to skip the computation of the zero values, we must expose some information about the structure of the sparsity in the encrypted matrices.
In such schemes, only information about the sparsity pattern is exposed with the matrix values remaining encrypted. We rationalize that this is acceptable for most privacy-sensitive DNN inference use cases; however, these sparse schemes may not be applicable in situations with very strict privacy requirements, such as DNNs that utilize one-hot encoding inputs, since the sparsity structure is enough information to reconstruct private DNN inputs.

\algrenewcommand{\algorithmicindent}{1em} 

\begin{algorithm}[ht]
\caption{Algorithm for naive dense matrix multiplication in FHE.}
\label{alg:dense-fhe-matmul}
\begin{algorithmic}[1]  
\footnotesize  

    \State \textbf{Input:} Encrypted list of chunks $A$ and $B$ representing respective matrices, and corresponding matrix information $LHS$ and $RHS$
    \State \textbf{Output:} Encrypted matrix product, stored in list of chunks, $R$
    
    \For{$row \gets 0$ to $|result_{\mathrm{rows}}| - 1$}  \Comment{\emph{Iterate over result rows}}
        \For{$col \gets 0$ to $|result_{\mathrm{cols}}| - 1$}  \Comment{\emph{Iterate over result columns}}
            \For{$k \gets 0$ to $|LHS_{\mathrm{cols}}| - 1$}  \Comment{\emph{Iterate over shared dimension}}

                \State $\cdots$ \Comment{\emph{Load chunks, calculate value offsets within chunks}}
                
                \State $a \gets \text{fhe\_rotate}(A_{\mathrm{chunk}}, A_{\mathrm{offset}})$  \Comment{\emph{Load LHS val. into slot zero}}
                \State $b \gets \text{fhe\_rotate}(B_{\mathrm{chunk}}, B_{\mathrm{offset}})$  \Comment{\emph{Load RHS val. into slot zero}}
                \State $a \gets a \times b$  \Comment{\emph{Multiply operands}}
                \State $a \gets \text{fhe\_relin}(a)$  \Comment{\emph{Re-linearize}}
                \State $a \gets \text{fhe\_rescale}(a)$  \Comment{\emph{Switch modulus and rescale}}

                \State $a \gets a \times \text{zero\_mask}$  \Comment{\emph{Multiply with slot zero mask}}
                \State $a \gets \text{fhe\_relin}(a)$  \Comment{\emph{Re-linearize}}
                \State $a \gets \text{fhe\_rescale}(a)$  \Comment{\emph{Switch modulus and rescale}}

                \State $a \gets \text{fhe\_rotate}(a, -R_{\mathrm{offset}})$  \Comment{\emph{Rotate prod. to result offset slot}}
                \State $R_{\mathrm{chunk}} \gets \text{fhe\_add}(R_{\mathrm{chunk}}, a)$  \Comment{\emph{Accum. into result chunk}}

            \EndFor
        \EndFor
    \EndFor

\end{algorithmic}
\end{algorithm}

We can circumvent this restriction in the situation where exposing the sparsity structure information is not acceptable. Some scenarios (e.g., cloud computing) may have knowledge of the weight sparsity structure and process the encrypted DNN input provided by the user without knowledge of the underlying sparse structure. As the model weights are already known in plaintext to the server and the user input is fully secure (including any matrix meta-data), this does not expose any additional information and is equivalent to performing a dense-sparse \verb|matmul|. However, this approach does not fully exploit the underlying sparsity in both operands and could therefore yield increased inference time.

We introduce three schemes for sparse FHE matrix multiplication, adapted from three plaintext algorithms: a na\"ive sparse scheme, CSR~\cite{saad_2003}, and ELLPACK~\cite{ellpack_sparse}.  

\textbf{Na\"ive Sparse.} We evaluate a na\"ive sparse implementation as a simple counterpart to our na\"ive dense implementation. We encrypt input matrix $M \in \mathbb{R}^{m \times n}$, including zero values into ciphertexts according to the chunking parameters. At instantiation, a binary matrix $B$ is constructed in parallel, where $B_{i,j} = [M_{i,j} = 0] \text{ } \forall i \in \{1, \dots, m\}, j \in \{1, \dots, n\} $. Matrix $B$ is exposed in plaintext during matrix multiplication, and we skip the computation of elements where the following condition holds: $B^{LHS}_{row,k} \lor B^{RHS}_{k,col}$.

\textbf{Compressed Sparse Row (CSR).} We evaluate the CSR format, which encodes non-zero values in the encrypted domain, storing the locations of these non-zero values in plaintext in row index and column index arrays. We choose not to implement Compressed Sparse Column (CSC), as it is equivalent to performing CSR on the transpose of the input matrix and both of them have equal number of operations for matrix multiplication. One possible optimization we do not explore in this work is transposing the RHS operand, which would then allow for efficient access of columns in the RHS operand.

\textbf{ELLPACK}. For input matrix $M \in \mathbb{R}^{m \times n}$, we encrypt $j$ values per row where $j = \max(NZV(M_{i, 0..n})) \text{ } \forall i \in \{1, \dots, m\}$ and $NZV(x)$ returns the number of non-zero values in a row. For each row, we encrypt the non-zero values and zero-pad as needed to reach length $j$. A parallel matrix $C$ denotes the column numbers for the elements in the encrypted matrix.



\subsection{Multi-Threading}

We further accelerate computation by introducing multi-threading to both dense and sparse multiplication schemes. We allocate one thread per result value, utilizing all available threads in the CPU. Synchronization primitives are used to coordinate access to result chunks; small chunks allow for uncoordinated accesses, improving threading performance at the cost of a higher memory footprint. Formally, we delegate the calculation of $A_{rows} \times B_{cols}$ resultant matrix values by allocating the computation among a pool of $n$ threads based on the index of the resultant value; $(row \times col) \mod n$. While computation occurs in parallel, threads must synchronize writing to result ciphertexts, as multiple result values may share a given ciphertext when $chunk size > 1$. 
In future work, we will look into using the GPU for acceleration, which has demonstrated many orders of magnitude increased performance in primitive homomorphic operations~\cite{phantom_fhe}.

\section{Evaluation}

In this section, we discuss the insights gained by evaluating our proposed sparse methods against dense baselines on CPU hardware. We first describe the profiling procedure. Then we discuss the execution time and memory performance of the sparse schemes and how our approach scaled in a multi-threaded environment. After that, we look at scaling to larger matrices beyond trivial examples and investigate the accuracy of our methods. Finally, we show more experimental results with a discussion around notable and surprising observations.


\subsection{Profiling Procedure}
\label{sec:profiling-procedure}

We repeat the following procedure three times for each sparsity level to capture performance variances. We record the execution time (excluding context setup, encryption, and decryption) of each FHE matrix multiplication scheme and the memory consumption of the ciphertexts involved.

\begin{enumerate}

    \item Generate operand matrices of size $n \times n$ with $\lfloor s \cdot n \rfloor$ elements set to zero, where $s$ denotes the desired sparsity level. Values are sampled from a Gaussian distribution $\mathcal{N}(0,1)$, representing weights from a DNN initialized with this distribution~\cite{franchi_bursuc_aldea_dubuisson_bloch, pytorch-initialisaiton}.
    
    \item Perform dense \verb|matmul| with Eigen3~\cite{eigenweb}, a C++ library for efficient linear algebra operations, to validate correctness for all following computations.
    
    \item Perform all sparse \verb|matmul| operations in plaintext to validate algorithmic correctness.
    
    \item Perform na\"ive and HEMat \verb|matmul| in FHE for baseline FHE performance.
    
    \item Perform sparse \verb|matmul| in FHE for all sparse implementations.
    
    \item Verify that all plaintext and FHE result values are within $\epsilon = 10^{-3}$ of the ground-truth Eigen3 result.
\end{enumerate}


\subsection{Runtime performance}

All evaluations are conducted on an AMD EPYC 7V13 64-core CPU. According to \autoref{sec:profiling-procedure}, we profile while multiplying two square matrices with sizes $2^{3} \times 2^{3}$, as this is the smallest square matrix size that allows us to allocate one thread per result element with $64$ threads and satisfies HEMat's requirements discussed in \autoref{sec:fhe-scheme}.

\begin{figure}
    \centering
    \includegraphics[width=0.99\linewidth]{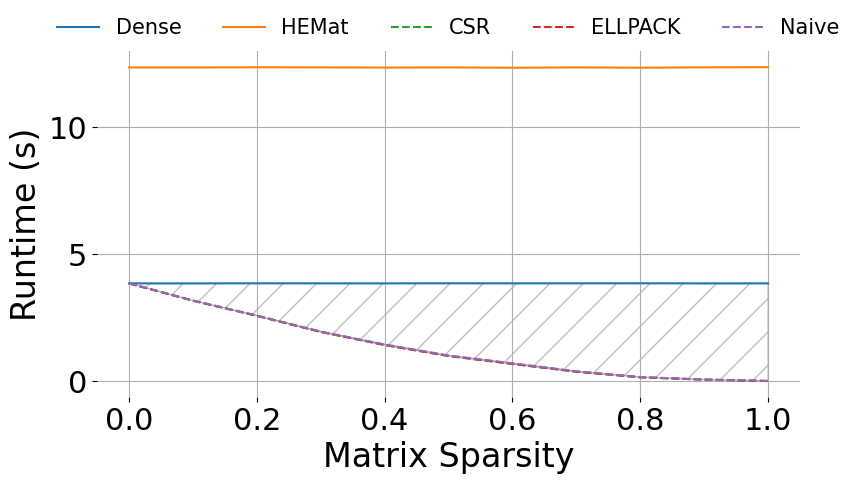}
    \caption{Runtime of dense and sparse schemes multiplying two $8 \times 8$ matrices. Shaded region denotes the runtime advantage of utilizing sparsity.}
    \label{fig:1_thread_8x8}
\end{figure}

\autoref{fig:1_thread_8x8} shows runtime performance when multiplying two $8 \times 8$ matrices on a single thread. As expected, both dense baseline algorithms that do not exploit any sparsity maintain a consistent runtime at varying sparsity levels. All sparse implementations display a runtime benefit over the dense schemes; moreover, we observe that the discrepancy between sparse schemes is negligible, with a standard deviation of runtime improvement of $0.039\times$ at 30\% sparsity. Furthermore, at this matrix size, our dense implementation runs $3.21\times$ faster than HEMat's on average.

In the plaintext domain, the ``break-even'' point (sparsity value where sparse \verb|matmul| becomes advantageous) has been observed to be $71\%$~\cite{gale_zaharia_brain_elsen_2020}. 
In our FHE implementation, it is at $0\%$ sparsity. We hypothesize that, while we incur additional overhead processing the sparse structures, it is shadowed by the computational burden of homomorphic operations and the large memory requirements of encrypted ciphertexts relative to the overhead of sparse data structures.


\subsection{Multi-Threading}

\begin{figure}[t]
    \centering
    \includegraphics[width=0.99\linewidth]{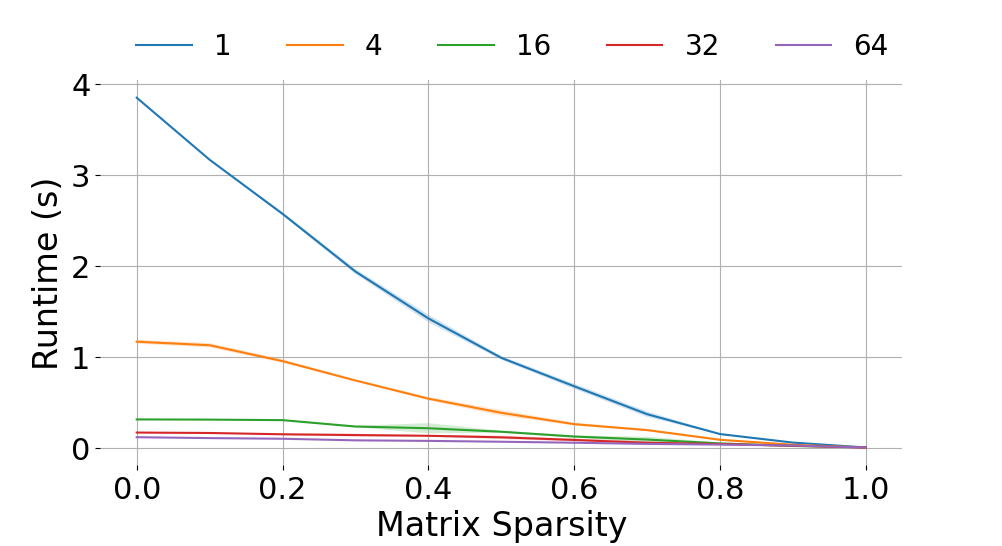}
    \caption{ELLPACK sparse scheme multi-threading \texttt{matmul} runtime. Multiplying $8 \times 8$ matrices with chunk size $c = 1$.}
    \label{fig:ellpack-runtime}
\end{figure}

Results from profiling performance on the ELLPACK sparse scheme can be found in \autoref{fig:ellpack-runtime} and \autoref{fig:ellpack-memory}. Our dense and sparse implementations observe similar gains with higher thread counts, with the ``break-even'' point remaining at 0\% sparsity. Results comparing the runtime of differing schemes at a given thread count can be found in \autoref{fig:1_thread_8x8}, \autoref{fig:threading_16_threads}, and \autoref{fig:threading_32_threads}. Notably, our implementation scales to higher thread counts more effectively with HEMat's threading scheme, which reaches diminishing returns at $16$ threads with a $3.46 \times$ performance gain compared to a single thread.

Our approach at $64$ threads exhibits a $32.47 \times$ performance gain from one thread in \autoref{fig:ellpack-runtime}. We experience diminishing returns with $64$ threads, providing a $1.39 \times$ gain compared to $32$. Additionally, we show negligible memory overhead in \autoref{fig:ellpack-memory}, only requiring synchronization primitives.

\subsection{Scaling to Larger Matrices}

Our analysis so far has been on $8 \times 8$ matrices. We now experiment with increasing matrix sizes and expect to scale to even higher dimensional matrices. Using $64$ threads, we evaluate the performance of \verb|matmul| at the following square matrix sizes: $8, 16, 32$. The results can be found in \autoref{fig:scaling_16x16} and \autoref{fig:scaling_32x32}, where we observe that our sparse schemes still exhibit a break-even point with the na\"ive dense implementation of $0\%$ and outperform HEMat at $\geq 10\%$ sparsity on matrix sizes of $32 \times 32$. 
However, our algorithms do not scale well to large matrix sizes relative to the HEMat implementation. We believe this is due to the poor $O(n^3)$ algorithmic complexity of our solution where HEMat exhibits $O(n)$ complexity; we leave scaling improvements for future work.


\subsection{Correctness}

We show the absolute mean error of each scheme at different sparsity levels in \autoref{tab:absolute_mean_error}. While we ensure that all schemes have a per-value error of less than $\epsilon = 10^{-3}$, accuracy differs between schemes. Our implementations provide a $7.6 \times 10^{-4}$ higher accuracy than the HEMat baseline, which may be advantageous in DNN models where precision is required. Furthermore, as sparsity trends to $100\%$, the sparse scheme results have no error as resultant zero values are effectively computed in plaintext.


\subsection{Discussion}

\begin{figure}[t]
    \centering
    \includegraphics[width=0.99\linewidth]{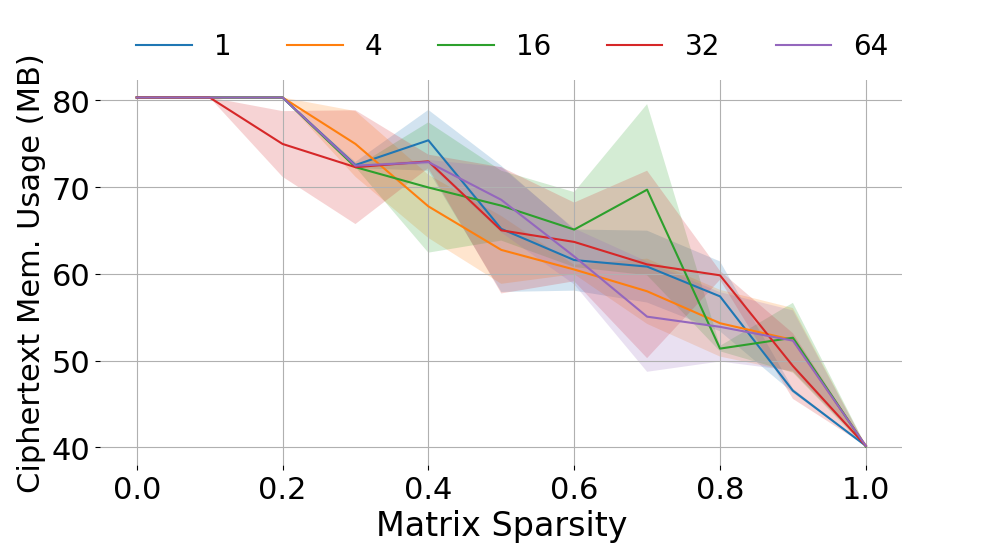}
    \caption{ELLPACK sparse scheme ciphertext memory usage. Multiplying $8 \times 8$ matrices with chunk size $c = 1$. The lines represent different thread count levels used for the computation with the shaded regions denoting $1 \sigma$ deviation, highlighting the low memory overhead of our multi-threading scheme relative to the ciphertext size.}
    \label{fig:ellpack-memory}
\end{figure}

We demonstrate that our proposed sparse methods require less memory to store operand ciphertexts in \autoref{fig:ellpack-memory}. This is demonstrated with the ELLPACK sparse scheme; however, similar memory savings are observed across all of our proposed sparse schemes.

\renewcommand{\arraystretch}{1.3}
\begin{table}[t]
\centering
\fontsize{9}{9}\selectfont
\caption{Mean absolute error between decrypted FHE matrices and plaintext ground truth results for multiplying $8 \times 8$ matrices with one thread and chunk size $c=1$. The lowest error scheme for each sparsity level is highlighted in \textbf{bold}. Our proposed sparse schemes exhibit higher accuracy as sparsity increases, a desirable attribute as we want encrypted computations to approximate plaintext inference closely. As we store sparsity information in plaintext, we compute zero values accurately and avoid noisy homomorphic operations. }
\label{tab:absolute_mean_error}
\begin{tabular}{l@{\hskip 4pt}c@{\hskip 4pt}c@{\hskip 4pt}c@{\hskip 4pt}c@{\hskip 4pt}c}
\toprule
\textbf{Sparsity} & \textbf{Dense} & \textbf{HEMat} & \textbf{CSR} & \textbf{ELLPACK} & \textbf{Na\"ive} \\ 
\midrule
0.0 & \textbf{5.98E-09} & 1.34E-03 & 6.03E-09 & 6.97E-09 & 6.66E-09 \\ 
0.1 & 5.63E-09 & 1.44E-03 & 5.74E-09 & 6.01E-09 & \textbf{5.14E-09} \\ 
0.2 & 5.47E-09 & 1.35E-03 & \textbf{5.15E-09} & 5.77E-09 & 6.44E-09 \\ 
0.3 & 6.03E-09 & 1.27E-03 & 6.29E-09 & \textbf{4.59E-09} & 5.11E-09 \\ 
0.4 & 4.84E-09 & 6.81E-04 & 3.45E-09 & \textbf{3.21E-09} & 3.38E-09 \\ 
0.5 & 5.75E-09 & 7.64E-04 & \textbf{3.24E-09} & 4.36E-09 & 3.55E-09 \\ 
0.6 & 5.06E-09 & 5.02E-04 & 2.72E-09 & \textbf{2.24E-09} & 2.80E-09 \\ 
0.7 & 4.73E-09 & 3.96E-04 & 1.38E-09 & \textbf{1.32E-09} & 1.77E-09 \\ 
0.8 & 4.84E-09 & 1.96E-04 & \textbf{5.44E-10} & 5.92E-10 & 1.31E-09 \\ 
0.9 & 4.61E-09 & 8.80E-05 & \textbf{2.85E-10} & 2.91E-10 & 1.28E-09 \\ 
1.0 & 4.40E-09 & 3.15E-06 & \textbf{0.00E+00} & 0.00E+00 & 8.90E-10 \\ 
\bottomrule
\end{tabular}
\end{table}

\begin{figure}[t]
    \centering
    \includegraphics[width=1.0\linewidth]{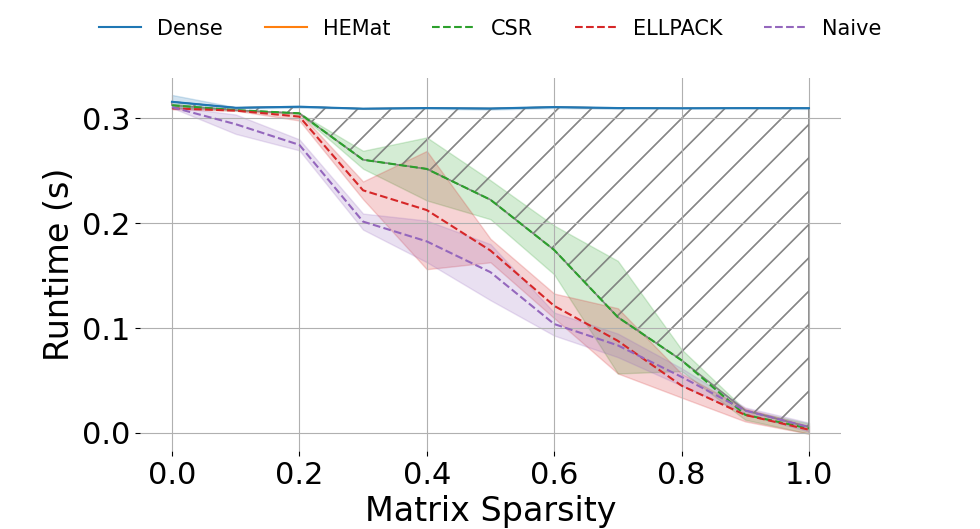}
    \caption{Relative runtime (HEMat excluded due to high runtime) between schemes when multiplying $8 \times 8$ matrices with 16 threads.}
    \label{fig:threading_16_threads}
\end{figure}

\begin{figure}[t]
    \centering
    \includegraphics[width=1.0\linewidth]{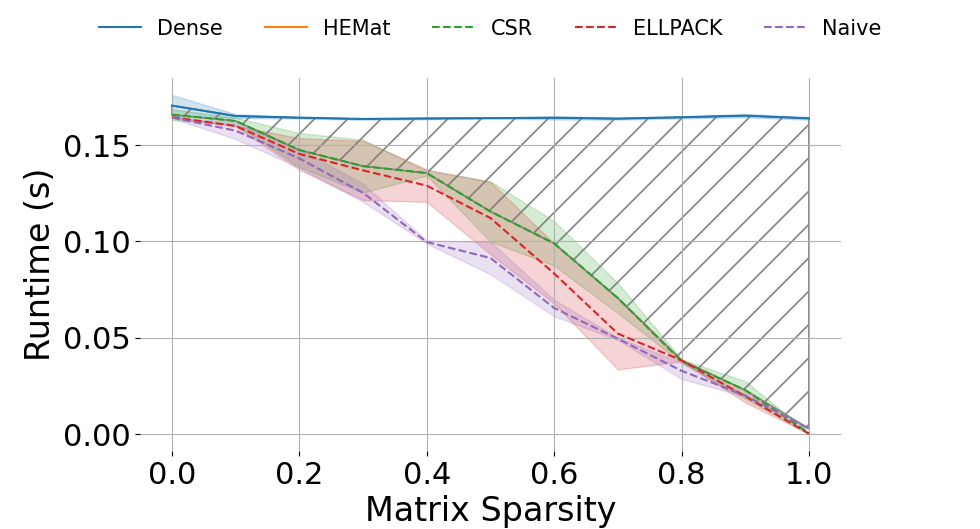}
    \caption{Relative runtime (HEMat excluded due to high runtime) between schemes when multiplying $8 \times 8$ matrices with 32 threads.}
    \label{fig:threading_32_threads}
\end{figure}

The runtime of our proposed sparse schemes as we utilize multi-threading on an AMD EPYC 7V13 64-core CPU can be seen in \autoref{fig:threading_16_threads} and \autoref{fig:threading_32_threads}. Our sparse schemes (denoted by dashed lines) are compared to our na\"ive implementation and HEMat baselines. Notably, our proposed schemes scale better in performance with more threads than HEMat, as such we remove the HEMat baseline from 16 threads and higher due to the difference between our methods and HEMat being too large for meaningful visualization. 
At $32$ threads, our multi-threading scheme exhibits over a $25 \times$ increase in speed. However, at thread counts beyond this, we reach diminishing returns, with a $32.5 \times$ performance increase at $64$ threads. 
Furthermore, our sparse schemes maintain a performance gain over the na\"ive baseline at all sparsity levels as the thread count varies.


\begin{figure}[t]
    \centering
    \includegraphics[width=0.91\linewidth]{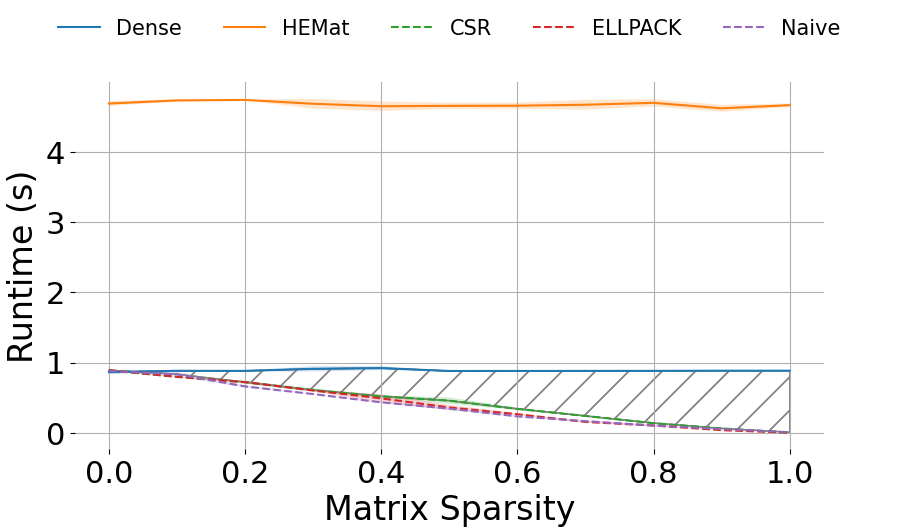}
    \caption{Relative runtime between multiplication schemes when multiplying $16 \times 16$ matrices with 64 threads.}
    \label{fig:scaling_16x16}
\end{figure}

\begin{figure}[t]
\vspace{5.5mm}
    \centering
    \includegraphics[width=0.91\linewidth]{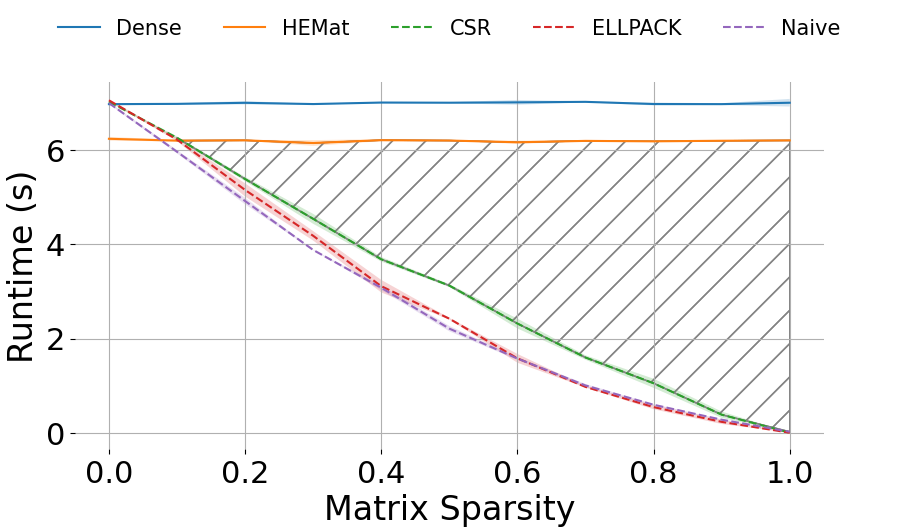}
    \caption{Relative runtime between multiplication schemes when multiplying $32 \times 32$ matrices with 64 threads. HEMat's superior algorithmic complexity has moved the ``break-even'' point to $\approx 0.1$ sparsity for this matrix size and thread count.}
    \label{fig:scaling_32x32}
\end{figure}

While we have demonstrated sparse matrix multiplication schemes in FHE that show improvements compared to baseline na\"ive performance, state-of-the-art solutions such as HEMat exhibit superior algorithmic complexity. These solutions scale better with respect to matrix size, a drawback that our approach exhibits with a $O(n^3)$ complexity. However, in many real-world situations, our approach outperforms these baselines as demonstrated in \autoref{fig:scaling_16x16} and \autoref{fig:scaling_32x32}. 

In summary, for small matrix sizes and high thread counts our solution outperforms HEMat, taking advantage of our multi-threading scheme. However, for larger matrix sizes, such as $32 \times 32$, HEMat is superior. The ``break-even'' sparsity threshold at which our proposed schemes are faster, increases. At our maximum evaluated thread count of $64$ threads, this begins to occur at matrix sizes of $32 \times 32$ and larger. However, for low thread counts this will happen at lower matrix sizes.

\section{Conclusion}

We have proposed sparse FHE matrix multiplication schemes within the context of DNN inference. We demonstrated a performance increase of $2.5 \times$ over a dense baseline implementation when operating at $50\%$ sparsity. Similarly, we show how our parallelism scheme can significantly increase performance with an observed $32.47 \times$ improvement over dense baselines on an AMD EPYC 7V13 64-core CPU. 

In future work, we aim to i) explore reducing the memory overhead of sparse matrix multiplication in GPU-accelerated FHE, ii) extend the use of sparsity beyond the $O(n^3)$ algorithmic approach and analyze end-to-end sparse inference GPU-accelerated FHE and iii) demonstrate the utilization of sparsity with larger matrices, such as those found in common small language models (SLMs) such as Pythia-1B~\cite{biderman2023pythia} that contain projection layers of size $4096 \times 1024$.


\begin{acks}
This work was supported in part by Advanced Micro Devices, Inc. under the AMD AI \& HPC Cluster Program.
\end{acks}

\newpage

\bibliographystyle{ACM-Reference-Format}
\bibliography{refs}


\begin{thebibliography}{40}


\ifx \showCODEN    \undefined \def \showCODEN     #1{\unskip}     \fi
\ifx \showDOI      \undefined \def \showDOI       #1{#1}\fi
\ifx \showISBNx    \undefined \def \showISBNx     #1{\unskip}     \fi
\ifx \showISBNxiii \undefined \def \showISBNxiii  #1{\unskip}     \fi
\ifx \showISSN     \undefined \def \showISSN      #1{\unskip}     \fi
\ifx \showLCCN     \undefined \def \showLCCN      #1{\unskip}     \fi
\ifx \shownote     \undefined \def \shownote      #1{#1}          \fi
\ifx \showarticletitle \undefined \def \showarticletitle #1{#1}   \fi
\ifx \showURL      \undefined \def \showURL       {\relax}        \fi
\providecommand\bibfield[2]{#2}
\providecommand\bibinfo[2]{#2}
\providecommand\natexlab[1]{#1}
\providecommand\showeprint[2][]{arXiv:#2}

\bibitem[Albrecht et~al\mbox{.}(2018)]%
        {HomomorphicEncryptionSecurityStandard}
\bibfield{author}{\bibinfo{person}{Martin Albrecht}, \bibinfo{person}{Melissa Chase}, \bibinfo{person}{Hao Chen}, \bibinfo{person}{Jintai Ding}, \bibinfo{person}{Shafi Goldwasser}, \bibinfo{person}{Sergey Gorbunov}, \bibinfo{person}{Shai Halevi}, \bibinfo{person}{Jeffrey Hoffstein}, \bibinfo{person}{Kim Laine}, \bibinfo{person}{Kristin Lauter}, \bibinfo{person}{Satya Lokam}, \bibinfo{person}{Daniele Micciancio}, \bibinfo{person}{Dustin Moody}, \bibinfo{person}{Travis Morrison}, \bibinfo{person}{Amit Sahai}, {and} \bibinfo{person}{Vinod Vaikuntanathan}.} \bibinfo{year}{2018}\natexlab{}.
\newblock \bibinfo{booktitle}{\emph{Homomorphic Encryption Security Standard}}.
\newblock \bibinfo{type}{{T}echnical {R}eport}. \bibinfo{institution}{HomomorphicEncryption.org}, \bibinfo{address}{Toronto, Canada}.
\newblock


\bibitem[Alexandru et~al\mbox{.}(2024)]%
        {alexandru2024general}
\bibfield{author}{\bibinfo{person}{Andreea Alexandru}, \bibinfo{person}{Andrey Kim}, {and} \bibinfo{person}{Yuriy Polyakov}.} \bibinfo{year}{2024}\natexlab{}.
\newblock \showarticletitle{General functional bootstrapping using CKKS}.
\newblock \bibinfo{journal}{\emph{Cryptology ePrint Archive}} (\bibinfo{year}{2024}).
\newblock


\bibitem[Benaissa et~al\mbox{.}(2021)]%
        {tenseal2021}
\bibfield{author}{\bibinfo{person}{Ayoub Benaissa}, \bibinfo{person}{Bilal Retiat}, \bibinfo{person}{Bogdan Cebere}, {and} \bibinfo{person}{Alaa~Eddine Belfedhal}.} \bibinfo{year}{2021}\natexlab{}.
\newblock \bibinfo{title}{TenSEAL: A Library for Encrypted Tensor Operations Using Homomorphic Encryption}.
\newblock
\newblock
\showeprint{2104.03152}~[cs.CR]


\bibitem[Biderman et~al\mbox{.}(2023)]%
        {biderman2023pythia}
\bibfield{author}{\bibinfo{person}{Stella Biderman}, \bibinfo{person}{Hailey Schoelkopf}, \bibinfo{person}{Quentin~Gregory Anthony}, \bibinfo{person}{Herbie Bradley}, \bibinfo{person}{Kyle O’Brien}, \bibinfo{person}{Eric Hallahan}, \bibinfo{person}{Mohammad~Aflah Khan}, \bibinfo{person}{Shivanshu Purohit}, \bibinfo{person}{USVSN~Sai Prashanth}, \bibinfo{person}{Edward Raff}, {et~al\mbox{.}}} \bibinfo{year}{2023}\natexlab{}.
\newblock \showarticletitle{Pythia: A suite for analyzing large language models across training and scaling}. In \bibinfo{booktitle}{\emph{International Conference on Machine Learning}}. PMLR, \bibinfo{pages}{2397--2430}.
\newblock


\bibitem[Bossuat et~al\mbox{.}(2021)]%
        {key-switching-costly}
\bibfield{author}{\bibinfo{person}{Jean-Philippe Bossuat}, \bibinfo{person}{Christian Mouchet}, \bibinfo{person}{Juan Troncoso-Pastoriza}, {and} \bibinfo{person}{Jean-Pierre Hubaux}.} \bibinfo{year}{2021}\natexlab{}.
\newblock \showarticletitle{Efficient Bootstrapping for Approximate Homomorphic Encryption with Non-sparse Keys}.
\newblock \bibinfo{journal}{\emph{Lecture notes in computer science}} (\bibinfo{date}{Jan} \bibinfo{year}{2021}), \bibinfo{pages}{587–617}.
\newblock
\urldef\tempurl%
\url{https://doi.org/10.1007/978-3-030-77870-5_21}
\showDOI{\tempurl}


\bibitem[Brakerski et~al\mbox{.}(2014)]%
        {bfv_citation}
\bibfield{author}{\bibinfo{person}{Zvika Brakerski}, \bibinfo{person}{Craig Gentry}, {and} \bibinfo{person}{Vinod Vaikuntanathan}.} \bibinfo{year}{2014}\natexlab{}.
\newblock \showarticletitle{(Leveled) Fully Homomorphic Encryption without Bootstrapping}.
\newblock \bibinfo{journal}{\emph{ACM Transactions on Computation Theory}} \bibinfo{volume}{6}, \bibinfo{number}{3} (\bibinfo{date}{Jul} \bibinfo{year}{2014}), \bibinfo{pages}{1–36}.
\newblock
\urldef\tempurl%
\url{https://doi.org/10.1145/2633600}
\showDOI{\tempurl}


\bibitem[Brakerski and Vaikuntanathan(2011)]%
        {bgv_citation}
\bibfield{author}{\bibinfo{person}{Zvika Brakerski} {and} \bibinfo{person}{Vinod Vaikuntanathan}.} \bibinfo{year}{2011}\natexlab{}.
\newblock \showarticletitle{Fully Homomorphic Encryption from Ring-LWE and Security for Key Dependent Messages}.
\newblock \bibinfo{journal}{\emph{Advances in Cryptology – CRYPTO 2011}} (\bibinfo{year}{2011}), \bibinfo{pages}{505–524}.
\newblock
\showISBNx{9783642227912}
\urldef\tempurl%
\url{https://doi.org/10.1007/978-3-642-22792-9_29}
\showDOI{\tempurl}


\bibitem[Chen et~al\mbox{.}(2021)]%
        {chen_zhou}
\bibfield{author}{\bibinfo{person}{Chaochao Chen}, \bibinfo{person}{Jun Zhou}, \bibinfo{person}{Li Wang}, \bibinfo{person}{Xibin Wu}, \bibinfo{person}{Wenjing Fang}, \bibinfo{person}{Jin Tan}, \bibinfo{person}{Lei Wang}, \bibinfo{person}{Alex~X Liu}, \bibinfo{person}{Hao Wang}, {and} \bibinfo{person}{Cheng~Shong Hong}.} \bibinfo{year}{2021}\natexlab{}.
\newblock \showarticletitle{When Homomorphic Encryption Marries Secret Sharing: Secure Large-Scale Sparse Logistic Regression and Applications in Risk Control}.
\newblock  (\bibinfo{date}{Aug} \bibinfo{year}{2021}).
\newblock
\urldef\tempurl%
\url{https://doi.org/10.1145/3447548.3467210}
\showDOI{\tempurl}


\bibitem[Chen et~al\mbox{.}(2015)]%
        {xiaofeng}
\bibfield{author}{\bibinfo{person}{Xiaofeng Chen}, \bibinfo{person}{Xinyi Huang}, \bibinfo{person}{Jin Li}, \bibinfo{person}{Jianfeng Ma}, \bibinfo{person}{Wenjing Lou}, {and} \bibinfo{person}{Duncan~S. Wong}.} \bibinfo{year}{2015}\natexlab{}.
\newblock \showarticletitle{New Algorithms for Secure Outsourcing of Large-Scale Systems of Linear Equations}.
\newblock \bibinfo{journal}{\emph{IEEE Transactions on Information Forensics and Security}} \bibinfo{volume}{10}, \bibinfo{number}{1} (\bibinfo{date}{Jan} \bibinfo{year}{2015}), \bibinfo{pages}{69–78}.
\newblock
\urldef\tempurl%
\url{https://doi.org/10.1109/tifs.2014.2363765}
\showDOI{\tempurl}


\bibitem[Cheon et~al\mbox{.}(2017)]%
        {cheon_kim_kim_song_2017}
\bibfield{author}{\bibinfo{person}{Jung~Hee Cheon}, \bibinfo{person}{Andrey Kim}, \bibinfo{person}{Miran Kim}, {and} \bibinfo{person}{Yongsoo Song}.} \bibinfo{year}{2017}\natexlab{}.
\newblock \showarticletitle{Homomorphic Encryption for Arithmetic of Approximate Numbers}.
\newblock \bibinfo{journal}{\emph{Advances in Cryptology – ASIACRYPT 2017}} (\bibinfo{year}{2017}), \bibinfo{pages}{409–437}.
\newblock
\showISBNx{9783319706931}
\urldef\tempurl%
\url{https://doi.org/10.1007/978-3-319-70694-8_15}
\showDOI{\tempurl}


\bibitem[Cui et~al\mbox{.}(2021)]%
        {cui_chen_lyu_yang_wang_2021}
\bibfield{author}{\bibinfo{person}{Jamie Cui}, \bibinfo{person}{Chaochao Chen}, \bibinfo{person}{Lingjuan Lyu}, \bibinfo{person}{Carl Yang}, {and} \bibinfo{person}{Li Wang}.} \bibinfo{year}{2021}\natexlab{}.
\newblock \bibinfo{booktitle}{\emph{Exploiting Data Sparsity in Secure Cross-Platform Social Recommendation}}.
\newblock
\urldef\tempurl%
\url{https://www.cs.emory.edu/~jyang71/files/s3rec.pdf}
\showURL{%
\tempurl}


\bibitem[Franchi et~al\mbox{.}({[n.\,d.]})]%
        {franchi_bursuc_aldea_dubuisson_bloch}
\bibfield{author}{\bibinfo{person}{Gianni Franchi}, \bibinfo{person}{Andrei Bursuc}, \bibinfo{person}{Emanuel Aldea}, \bibinfo{person}{Séverine Dubuisson}, {and} \bibinfo{person}{Isabelle Bloch}.} \bibinfo{year}{[n.\,d.]}\natexlab{}.
\newblock \bibinfo{booktitle}{\emph{TRADI: Tracking deep neural network weight distributions}}.
\newblock
\urldef\tempurl%
\url{https://www.ecva.net/papers/eccv_2020/papers_ECCV/papers/123620103.pdf}
\showURL{%
\tempurl}


\bibitem[Gale et~al\mbox{.}(2020)]%
        {gale_zaharia_brain_elsen_2020}
\bibfield{author}{\bibinfo{person}{Trevor Gale}, \bibinfo{person}{Matei Zaharia}, \bibinfo{person}{Cliff Young}, {and} \bibinfo{person}{Erich Elsen}.} \bibinfo{year}{2020}\natexlab{}.
\newblock \showarticletitle{Sparse GPU kernels for deep learning}. In \bibinfo{booktitle}{\emph{Proceedings of the International Conference for High Performance Computing, Networking, Storage and Analysis}} (Atlanta, Georgia) \emph{(\bibinfo{series}{SC '20})}. \bibinfo{publisher}{IEEE Press}, Article \bibinfo{articleno}{17}, \bibinfo{numpages}{14}~pages.
\newblock
\showISBNx{9781728199986}


\bibitem[Gentry(2009)]%
        {gentry_2009}
\bibfield{author}{\bibinfo{person}{Craig Gentry}.} \bibinfo{year}{2009}\natexlab{}.
\newblock \showarticletitle{Fully homomorphic encryption using ideal lattices}.
\newblock \bibinfo{journal}{\emph{Proceedings of the 41st annual ACM symposium on Symposium on theory of computing - STOC ’09}} (\bibinfo{year}{2009}).
\newblock
\showISBNx{9781605585062}
\urldef\tempurl%
\url{https://doi.org/10.1145/1536414.1536440}
\showDOI{\tempurl}


\bibitem[Guennebaud et~al\mbox{.}(2010)]%
        {eigenweb}
\bibfield{author}{\bibinfo{person}{Ga\"{e}l Guennebaud}, \bibinfo{person}{Beno\^{i}t Jacob}, {et~al\mbox{.}}} \bibinfo{year}{2010}\natexlab{}.
\newblock \bibinfo{title}{Eigen v3}.
\newblock \bibinfo{howpublished}{http://eigen.tuxfamily.org}.
\newblock


\bibitem[Gupta et~al\mbox{.}(2022)]%
        {gupta_gupta_shabaz_sharma_2022}
\bibfield{author}{\bibinfo{person}{Surbhi Gupta}, \bibinfo{person}{Manoj~K. Gupta}, \bibinfo{person}{Mohammad Shabaz}, {and} \bibinfo{person}{Ashutosh Sharma}.} \bibinfo{year}{2022}\natexlab{}.
\newblock \showarticletitle{Deep learning techniques for cancer classification using microarray gene expression data}.
\newblock \bibinfo{journal}{\emph{Frontiers in Physiology}}  \bibinfo{volume}{13} (\bibinfo{date}{Sep} \bibinfo{year}{2022}).
\newblock
\urldef\tempurl%
\url{https://doi.org/10.3389/fphys.2022.952709}
\showDOI{\tempurl}


\bibitem[Haris et~al\mbox{.}(2021)]%
        {haris_SECDA_2021}
\bibfield{author}{\bibinfo{person}{Jude Haris}, \bibinfo{person}{Perry Gibson}, \bibinfo{person}{Jos{\'e} Cano}, \bibinfo{person}{Nicolas~Bohm Agostini}, {and} \bibinfo{person}{David Kaeli}.} \bibinfo{year}{2021}\natexlab{}.
\newblock \showarticletitle{{{SECDA}}: {{Efficient Hardware}}/{{Software Co-Design}} of {{FPGA-based DNN Accelerators}} for {{Edge Inference}}}. In \bibinfo{booktitle}{\emph{2021 {{IEEE}} 33rd {{International Symposium}} on {{Computer Architecture}} and {{High Performance Computing}} ({{SBAC-PAD}})}}.
\newblock


\bibitem[Haris et~al\mbox{.}(2023)]%
        {haris_SECDA-TFLite_2023}
\bibfield{author}{\bibinfo{person}{Jude Haris}, \bibinfo{person}{Perry Gibson}, \bibinfo{person}{José Cano}, \bibinfo{person}{Nicolas Bohm~Agostini}, {and} \bibinfo{person}{David Kaeli}.} \bibinfo{year}{2023}\natexlab{}.
\newblock \showarticletitle{{{SECDA-TFLite}}: {{A}} Toolkit for Efficient Development of {{FPGA-based DNN}} Accelerators for Edge Inference}. In \bibinfo{booktitle}{\emph{{Journal of Parallel and Distributed Computing}}}.
\newblock


\bibitem[Haris et~al\mbox{.}(2024)]%
        {haris_SECDA-LLM_2024}
\bibfield{author}{\bibinfo{person}{Jude Haris}, \bibinfo{person}{Rappy Saha}, \bibinfo{person}{Wenhao Hu}, {and} \bibinfo{person}{José Cano}.} \bibinfo{year}{2024}\natexlab{}.
\newblock \showarticletitle{{Designing Efficient LLM Accelerators for Edge Devices}}. In \bibinfo{booktitle}{\emph{{Workshop on New Approaches for Addressing the Computing Requirements of LLMs and GNNs (ARC-LG) at ISCA}}}.
\newblock


\bibitem[He et~al\mbox{.}(2015)]%
        {He2015DeepRL}
\bibfield{author}{\bibinfo{person}{Kaiming He}, \bibinfo{person}{X. Zhang}, \bibinfo{person}{Shaoqing Ren}, {and} \bibinfo{person}{Jian Sun}.} \bibinfo{year}{2015}\natexlab{}.
\newblock \showarticletitle{Deep Residual Learning for Image Recognition}.
\newblock \bibinfo{journal}{\emph{2016 IEEE Conference on Computer Vision and Pattern Recognition (CVPR)}}, \bibinfo{pages}{770--778}.
\newblock


\bibitem[Hiromasa et~al\mbox{.}(2016)]%
        {hiromasa_abe_okamoto_2016}
\bibfield{author}{\bibinfo{person}{Ryo Hiromasa}, \bibinfo{person}{Masayuki Abe}, {and} \bibinfo{person}{Tatsuaki Okamoto}.} \bibinfo{year}{2016}\natexlab{}.
\newblock \showarticletitle{Packing messages and optimizing bootstrapping in GSW-FHE}.
\newblock \bibinfo{journal}{\emph{IEICE TRANSACTIONS on Fundamentals of Electronics, Communications and Computer Sciences}} \bibinfo{volume}{99}, \bibinfo{number}{1} (\bibinfo{year}{2016}), \bibinfo{pages}{73--82}.
\newblock


\bibitem[Hu et~al\mbox{.}(2016)]%
        {network_trimming}
\bibfield{author}{\bibinfo{person}{Hengyuan Hu}, \bibinfo{person}{Rui Peng}, \bibinfo{person}{Yu-Wing Tai}, {and} \bibinfo{person}{Chi-Keung Tang}.} \bibinfo{year}{2016}\natexlab{}.
\newblock \showarticletitle{Network Trimming: A Data-Driven Neuron Pruning Approach towards Efficient Deep Architectures}.
\newblock \bibinfo{journal}{\emph{arXiv:1607.03250 [cs]}} (\bibinfo{date}{Jul} \bibinfo{year}{2016}).
\newblock
\urldef\tempurl%
\url{https://arxiv.org/abs/1607.03250}
\showURL{%
\tempurl}


\bibitem[Huang and Zong(2022)]%
        {huang_zong_2022}
\bibfield{author}{\bibinfo{person}{Hai Huang} {and} \bibinfo{person}{Haoran Zong}.} \bibinfo{year}{2022}\natexlab{}.
\newblock \showarticletitle{Secure matrix multiplication based on fully homomorphic encryption}.
\newblock \bibinfo{journal}{\emph{The Journal of Supercomputing}} (\bibinfo{date}{Oct} \bibinfo{year}{2022}).
\newblock
\urldef\tempurl%
\url{https://doi.org/10.1007/s11227-022-04850-4}
\showDOI{\tempurl}


\bibitem[Jiang et~al\mbox{.}(2023)]%
        {jiang_zhong_zhou_2023}
\bibfield{author}{\bibinfo{person}{Jiachen Jiang}, \bibinfo{person}{Yiqi Zhong}, {and} \bibinfo{person}{Jinxin Zhou}.} \bibinfo{year}{2023}\natexlab{}.
\newblock \bibinfo{booktitle}{\emph{The Efficiency Spectrum of Large Language Models: An Algorithmic Survey}}.
\newblock
\urldef\tempurl%
\url{https://arxiv.org/pdf/2312.00678}
\showURL{%
\tempurl}


\bibitem[Jiang et~al\mbox{.}(2018)]%
        {hemat}
\bibfield{author}{\bibinfo{person}{Xiaoqian Jiang}, \bibinfo{person}{Miran Kim}, \bibinfo{person}{Kristin~E Lauter}, {and} \bibinfo{person}{Yongsoo Song}.} \bibinfo{year}{2018}\natexlab{}.
\newblock \showarticletitle{Secure Outsourced Matrix Computation and Application to Neural Networks}.
\newblock  (\bibinfo{date}{Oct} \bibinfo{year}{2018}).
\newblock
\urldef\tempurl%
\url{https://doi.org/10.1145/3243734.3243837}
\showDOI{\tempurl}


\bibitem[K-miran(2018)]%
        {hemat_github}
\bibfield{author}{\bibinfo{person}{K-miran}.} \bibinfo{year}{2018}\natexlab{}.
\newblock \bibinfo{title}{GitHub - K-miran/HEMat: Homomorphic matrix computation}.
\newblock
\newblock
\urldef\tempurl%
\url{https://github.com/K-miran/HEMat}
\showURL{%
\tempurl}


\bibitem[Kincaid et~al\mbox{.}(1989)]%
        {ellpack_sparse}
\bibfield{author}{\bibinfo{person}{D Kincaid}, \bibinfo{person}{T Oppe}, {and} \bibinfo{person}{D Young}.} \bibinfo{year}{1989}\natexlab{}.
\newblock \showarticletitle{ITPACKV 2D user’s guide}.
\newblock \bibinfo{journal}{\emph{OSTI OAI (U.S. Department of Energy Office of Scientific and Technical Information)}} (\bibinfo{date}{May} \bibinfo{year}{1989}).
\newblock
\urldef\tempurl%
\url{https://doi.org/10.2172/7093021}
\showDOI{\tempurl}


\bibitem[Lee et~al\mbox{.}(2023)]%
        {relu_approximation}
\bibfield{author}{\bibinfo{person}{Junghyun Lee}, \bibinfo{person}{Eunsang Lee}, \bibinfo{person}{Joon-Woo Lee}, \bibinfo{person}{Yongjune Kim}, \bibinfo{person}{Young-Sik Kim}, {and} \bibinfo{person}{Jong-Seon No}.} \bibinfo{year}{2023}\natexlab{}.
\newblock \showarticletitle{Precise Approximation of Convolutional Neural Networks for Homomorphically Encrypted Data}.
\newblock \bibinfo{journal}{\emph{IEEE Access}}  \bibinfo{volume}{11} (\bibinfo{date}{06} \bibinfo{year}{2023}), \bibinfo{pages}{62062 -- 62076}.
\newblock
\urldef\tempurl%
\url{https://doi.org/10.1109/ACCESS.2023.3287564}
\showDOI{\tempurl}


\bibitem[Lee et~al\mbox{.}(2022)]%
        {privacy_preserving_ml}
\bibfield{author}{\bibinfo{person}{Joon-Woo Lee}, \bibinfo{person}{Hyungchul Kang}, \bibinfo{person}{Yongwoo Lee}, \bibinfo{person}{Woosuk Choi}, \bibinfo{person}{Jieun Eom}, \bibinfo{person}{Maxim Deryabin}, \bibinfo{person}{Eunsang Lee}, \bibinfo{person}{Junghyun Lee}, \bibinfo{person}{Donghoon Yoo}, \bibinfo{person}{Young-Sik Kim}, {and} \bibinfo{person}{Jong-Seon No}.} \bibinfo{year}{2022}\natexlab{}.
\newblock \showarticletitle{Privacy-Preserving Machine Learning With Fully Homomorphic Encryption for Deep Neural Network}.
\newblock \bibinfo{journal}{\emph{IEEE Access}}  \bibinfo{volume}{10} (\bibinfo{year}{2022}), \bibinfo{pages}{30039--30054}.
\newblock
\urldef\tempurl%
\url{https://doi.org/10.1109/ACCESS.2022.3159694}
\showDOI{\tempurl}


\bibitem[Ma{\u{g}}ara et~al\mbox{.}(2021)]%
        {maugara2021ml}
\bibfield{author}{\bibinfo{person}{{\c{S}}S Ma{\u{g}}ara}, \bibinfo{person}{C Y{\i}ld{\i}r{\i}m}, \bibinfo{person}{F Yaman}, \bibinfo{person}{B Dileko{\u{g}}lu}, \bibinfo{person}{FR Tuta{\c{s}}}, \bibinfo{person}{E {\"O}zt{\"u}rk}, \bibinfo{person}{K Kaya}, \bibinfo{person}{{\"O} Ta{\c{s}}tan}, {and} \bibinfo{person}{E Sava{\c{s}}}.} \bibinfo{year}{2021}\natexlab{}.
\newblock \showarticletitle{ML with HE: Privacy Preserving Machine Learning Inferences for Genome Studies}.
\newblock \bibinfo{journal}{\emph{arXiv e-prints}} (\bibinfo{year}{2021}), \bibinfo{pages}{arXiv--2110}.
\newblock


\bibitem[Mirzadeh et~al\mbox{.}(2023)]%
        {relu_strikes_back}
\bibfield{author}{\bibinfo{person}{Iman Mirzadeh}, \bibinfo{person}{Keivan Alizadeh}, \bibinfo{person}{Sachin Mehta}, \bibinfo{person}{Carlo Del}, \bibinfo{person}{Mundo Oncel}, \bibinfo{person}{Tuzel Golnoosh}, \bibinfo{person}{Samei Mohammad}, \bibinfo{person}{Rastegari Mehrdad}, {and} \bibinfo{person}{Farajtabar Apple}.} \bibinfo{year}{2023}\natexlab{}.
\newblock \bibinfo{booktitle}{\emph{ReLU Strikes Back: Exploiting Activation Sparsity in Large Language Models}}.
\newblock
\urldef\tempurl%
\url{https://arxiv.org/pdf/2310.04564}
\showURL{%
\tempurl}


\bibitem[Onoufriou et~al\mbox{.}(2022)]%
        {ckks_justifications}
\bibfield{author}{\bibinfo{person}{George Onoufriou}, \bibinfo{person}{Marc Hanheide}, {and} \bibinfo{person}{Georgios Leontidis}.} \bibinfo{year}{2022}\natexlab{}.
\newblock \showarticletitle{EDLaaS: Fully Homomorphic Encryption over Neural Network Graphs for Vision and Private Strawberry Yield Forecasting}.
\newblock \bibinfo{journal}{\emph{Sensors}} \bibinfo{volume}{22}, \bibinfo{number}{21} (\bibinfo{year}{2022}).
\newblock
\showISSN{1424-8220}
\urldef\tempurl%
\url{https://doi.org/10.3390/s22218124}
\showDOI{\tempurl}


\bibitem[Pope et~al\mbox{.}(2022)]%
        {transformer_performance}
\bibfield{author}{\bibinfo{person}{Reiner Pope}, \bibinfo{person}{Sholto Douglas}, \bibinfo{person}{Aakanksha Chowdhery}, \bibinfo{person}{Jacob Devlin}, \bibinfo{person}{James Bradbury}, \bibinfo{person}{Anselm Levskaya}, \bibinfo{person}{Jonathan Heek}, \bibinfo{person}{Kefan Xiao}, \bibinfo{person}{Shivani Agrawal}, {and} \bibinfo{person}{Jeff Dean}.} \bibinfo{year}{2022}\natexlab{}.
\newblock \bibinfo{title}{Efficiently Scaling Transformer Inference}.
\newblock
\newblock
\urldef\tempurl%
\url{https://arxiv.org/abs/2211.05102}
\showURL{%
\tempurl}


\bibitem[PyTorch({[n.\,d.]})]%
        {pytorch-initialisaiton}
\bibfield{author}{\bibinfo{person}{PyTorch}.} \bibinfo{year}{[n.\,d.]}\natexlab{}.
\newblock \bibinfo{title}{torch.nn.init — PyTorch 2.2 documentation}.
\newblock
\newblock
\urldef\tempurl%
\url{https://pytorch.org/docs/stable/nn.init.html}
\showURL{%
\tempurl}


\bibitem[Rathee et~al\mbox{.}(2018)]%
        {deevashwer}
\bibfield{author}{\bibinfo{person}{Deevashwer Rathee}, \bibinfo{person}{Pradeep~Kumar Mishra}, {and} \bibinfo{person}{Masaya Yasuda}.} \bibinfo{year}{2018}\natexlab{}.
\newblock \showarticletitle{Faster PCA and Linear Regression through Hypercubes in HElib}.
\newblock \bibinfo{journal}{\emph{2018 Workshop on Privacy in the Electronic Society}} (\bibinfo{date}{Jan} \bibinfo{year}{2018}).
\newblock
\urldef\tempurl%
\url{https://doi.org/10.1145/3267323.3268952}
\showDOI{\tempurl}


\bibitem[Saad(2003)]%
        {saad_2003}
\bibfield{author}{\bibinfo{person}{Yousef Saad}.} \bibinfo{year}{2003}\natexlab{}.
\newblock \bibinfo{booktitle}{\emph{Iterative Methods for Sparse Linear Systems Second Edition Yousef Saad}}.
\newblock
\urldef\tempurl%
\url{https://www-users.cse.umn.edu/~saad/IterMethBook_2ndEd.pdf}
\showURL{%
\tempurl}


\bibitem[SEAL(2023)]%
        {sealcrypto}
SEAL \bibinfo{year}{2023}\natexlab{}.
\newblock \bibinfo{title}{{M}icrosoft {SEAL} (release 4.1)}.
\newblock \bibinfo{howpublished}{\url{https://github.com/Microsoft/SEAL}}.
\newblock
\newblock
\shownote{Microsoft Research, Redmond, WA}.


\bibitem[Simonyan and Zisserman(2015)]%
        {Simonyan15}
\bibfield{author}{\bibinfo{person}{Karen Simonyan} {and} \bibinfo{person}{Andrew Zisserman}.} \bibinfo{year}{2015}\natexlab{}.
\newblock \showarticletitle{Very Deep Convolutional Networks for Large-Scale Image Recognition}. In \bibinfo{booktitle}{\emph{International Conference on Learning Representations}}.
\newblock


\bibitem[snucrypto(2023)]%
        {snucrypto_2023}
\bibfield{author}{\bibinfo{person}{snucrypto}.} \bibinfo{year}{2023}\natexlab{}.
\newblock \bibinfo{title}{HEAAN}.
\newblock \bibinfo{howpublished}{\url{https://github.com/snucrypto/HEAAN}}.
\newblock
\urldef\tempurl%
\url{https://github.com/snucrypto/HEAAN}
\showURL{%
\tempurl}


\bibitem[Yang et~al\mbox{.}(2024)]%
        {phantom_fhe}
\bibfield{author}{\bibinfo{person}{Hao Yang}, \bibinfo{person}{Shiyu Shen}, \bibinfo{person}{Wangchen Dai}, \bibinfo{person}{Lu Zhou}, \bibinfo{person}{Zhe Liu}, {and} \bibinfo{person}{Yunlei Zhao}.} \bibinfo{year}{2024}\natexlab{}.
\newblock \showarticletitle{Phantom: A CUDA-Accelerated Word-Wise Homomorphic Encryption Library}.
\newblock \bibinfo{journal}{\emph{IEEE Transactions on Dependable and Secure Computing}} (\bibinfo{year}{2024}), \bibinfo{pages}{1--12}.
\newblock
\urldef\tempurl%
\url{https://doi.org/10.1109/TDSC.2024.3363900}
\showDOI{\tempurl}


\end{thebibliography}


\end{document}